\documentclass[a4paper,onecolumn,11pt,unpublished]{quantumarticle}
\pdfoutput=1
\usepackage[utf8]{inputenc}
\usepackage[english]{babel}
\usepackage[T1]{fontenc}
\usepackage{amsmath,amssymb}
\usepackage{hyperref}

\usepackage[squaren]{SIunits}

\newcommand{\eg}{\textit{e.g. }}
\newcommand{\ie}{\textit{i.e. }}
\newcommand{\etal}{\textit{et al. }}

\newcommand{\orho}{\hat{\rho}}
\newcommand{\oa}{\hat{a}}
\newcommand{\ox}{\hat{X}}
\newcommand{\op}{\hat{P}}
\newcommand{\vq}{\mathbf{q}}
\newcommand{\vu}{\mathbf{u}}
\newcommand{\ve}{\mathbf{e}}
\newcommand{\vr}{\mathbf{r}}

\newcommand{\vx}{\mathbf{x}}
\newcommand{\vy}{\mathbf{y}}
\newcommand{\vz}{\mathbf{z}}
\newcommand{\oM}{\hat{M}}
\newcommand{\vrop}{\hat{\mathbf{r}}}

\newcommand{\gammadown}{\gamma_{\downarrow}}

\begin{document}

\title{Probing gravity-related decoherence with a 16\,$\mu$g Schrödinger cat state}

\author{Matteo Fadel}
\affiliation{Department of Physics, ETH Z\"urich, 8093 Z\"urich, Switzerland}
\orcid{0000-0003-3653-0030}
\email{fadelm@phys.ethz.ch}

\begin{abstract}
    The Schrödinger equation predicts the validity of the superposition principle at any scale, yet we do not experience cats being in a superposition of ``dead'' and ``alive'' in our everyday lives. Modifications to quantum theory at the fundamental level may be responsible for the objective collapse of the wave function above a critical mass, thereby breaking down the superposition principle and restoring classical behavior at the macroscopic scale. One possibility is that these modifications are related to gravity, as described by the Diósi-Penrose wavefunction collapse model. Here, we investigate this model using experimental measurements on the decoherence of a Schrödinger cat state of a mechanical resonator with an effective mass of 16 micrograms.
\end{abstract}

\maketitle

\begin{figure}[h]
  \centering
  \includegraphics[width=\textwidth]{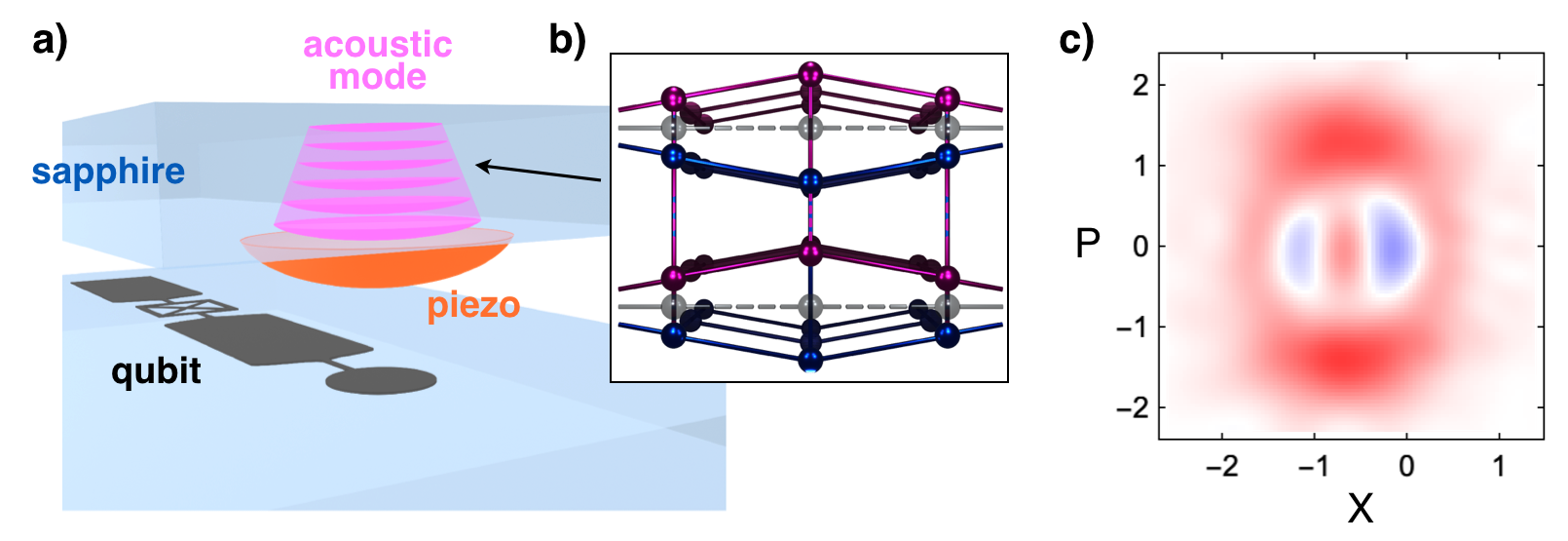}
  \caption{\textbf{Schrödinger cat state of a bulk acoustic resonator.} a) Illustration of the device used for the experiment, where a slab of piezoelectric material couples a superconducting qubit to the bulk acoustic mode of a sapphire crystal resonator \cite{catSCI23}. b) When the acoustic mode is prepared in a Schrödinger cat state of motion, the atoms of the crystal are in a superposition of oscillating with two opposite phases (not to scale). c) Reconstructed Wigner function of the Schrödinger cat state prepared in the resonator \cite{catSCI23}. Monitoring the decoherence of this state allows us to test the Diósi-Penrose gravity-related wavefunction collapse model.}
  \label{fig:figure1}
\end{figure}

\clearpage
\newpage

\section{Introduction}

The theory of quantum mechanics has proven to be an incredibly powerful and accurate framework for describing the behavior of fundamental particles and their interactions. However, despite its many successes, there remains a fundamental puzzle: why do we not observe quantum mechanical effects on arbitrarily large objects? 

One possible explanation is that the Schrödinger equation, which forms the backbone of quantum mechanics, is merely the approximation of a more fundamental theory that is still unknown. Consequently, researchers have proposed a variety of phenomenological modifications to this equation, designed to restore classical behavior at the macroscopic scale. These ``classicalizing'' modifications can take the form of nonlinear and/or stochastic terms, and be motivated by \eg gravitational effects \cite{diosi87,Penrose96,pikovski15,bassiGrav17}, the discretisation of space-time at the Planck scale \cite{petruzziello21}, or randomly fluctuating fields \cite{GRW86,GhirardiPearle90,BassiRMP13}.

The idea behind relating such a modification to gravity can be intuitively understood from the argument put forward by Penrose \cite{Penrose96}: a massive object in a quantum superposition of distinct locations must create a quantum superposition of spacetimes with a difference in gravitational energy $\Delta E$, which, according to the time-energy uncertainty principle, should dephase on a timescale $\tau \sim \hbar/\Delta E$.
Following different arguments, Diósi proposed a collapse model involving gravity by relating the postulated stochastic noise field to a classical Newtonian potential \cite{diosi87,diosi89}. Remarkably, this dynamical model reproduces Penrose's prediction and, for this reason, is commonly known in the literature as the Diósi-Penrose (DP) model.

In the last decade, the DP model has been the subject of a number of experimental tests. In particular, one advantage of this model, like other collapse models, is that it can be indirectly tested without the need to prepare massive superposition states or even quantum states at all. The reason being that the stochastic dynamics in the model predicts heating in bulk matter \cite{ghirardi90}, or the emission of photons due to the Brownian-like diffusion of its charged constituent particles \cite{donadi21}.

However, it is still of fundamental interest and conceptually different to perform a direct test of the DP model using a massive system in a spatial superposition state. This would show a genuine quantum effect and its decoherence dynamics, possibly demonstrating an incompatibility with standard quantum mechanics.
For this reason, here we perform such a test using the data of a solid-state mechanical resonator prepared in a Schrödinger cat state of motion that are presented in Ref.\cite{catSCI23}. The mechanical mode we consider is an acoustic excitation with an effective mass of $\unit{16.2}{\mu g}$, consisting of approximately $10^{17}$ atoms in a superposition of oscillating with two opposite phases, see Fig.~\ref{fig:figure1}.

\clearpage
\newpage
\section{The Diósi-Penrose model}

Formally, standard quantum mechanics postulates that the state $\orho$ of a system evolves in time as $\dot{\orho}=\mathcal{L}_\text{QM}[\orho]$, where $\mathcal{L}_\text{QM}$ is the Liouvillian describing both the unitary (von Neumann) evolution $[H,\orho]/i\hbar$, as well as potential non-unitary evolutions due to coupling with the environment. However, in the context of modified quantum mechanics, the time evolution of the state is phenomenologically described by $\dot{\orho}=\mathcal{L}_\text{QM}[\orho]+\mathcal{L}[\orho]$, with $\mathcal{L}$ being the modification term. Here we focus on the Diósi-Penrose (DP) model, where
\begin{equation}\label{eq:Lform}
    \mathcal{L}[\orho] = \dfrac{G}{\hbar} \int\int \dfrac{d^3\vx d^3\vy}{\vert \vx - \vy \vert} \left( \oM(\vx)\orho\oM(\vy) - \dfrac{1}{2}\{ \oM(\vx)\oM(\vy) , \orho \} \right) \;,
\end{equation}
with 
$G$ the universal gravitational constant, and $\oM$ the local mass density operator. This expression arises from a stochastic nonlinear Schrödinger equation in which the two-point correlation function of the collapse field is related to a Newtonian gravitational potential \cite{diosi87}, and remarkably it does not contain any free parameter. For a free particle, Eq.~\eqref{eq:Lform} predicts a decay of spatial superpositions at a rate $\hbar/\Delta E$, where $\Delta E$ is the difference in gravitational self-energy, in accordance to Penrose's estimate \cite{Penrose96}.

If one considers a point-like density operator $\oM(\vx)=\sum_{j=1}^N m_j \delta(\vx-\vrop_j)$, with $\vrop_j$ the position operator of the $j$ particle with mass $m_j$, then Eq.~\eqref{eq:Lform} gives diverging terms. For this reason, it was proposed to introduce a cutoff which regularizes the dynamics. This cutoff can be expressed in the form of a coarse-grained mass density operator \cite{ghirardi90}
\begin{equation}\label{eq:MDensOp}
    \oM(\vx) \rightarrow \int d\vz \, \dfrac{\exp[-|\vx-\vz|^2/2 R_0^2]}{(2\pi R_0^2)^{3/2}} \oM(\vz) \;,
\end{equation}
where $R_0$ defines the spatial resolution of the collapse field. In other words, it sets an upper bound $q_\text{max}=\hbar/R_0$ on the momentum of the field's modes contributing to the collapse. A smaller value of $R_0$ leads to decay of spatial superpositions that is faster, and with higher spatial resolution. Originally, $R_0$ was chosen to be equal to the Compton wavelength of a nucleon, $2\pi\hbar/mc\sim\unit{10^{-15}}{m}$, which is also consistent with the nonrelativistic framework in which the model applies. However, this is incompatible with experimental observations since, in the absence of other dissipative mechanisms, the collapse dynamics leas to an energy increase at rate $dE/dt\sim mG\hbar/R_0^3$, that for very small values of $R_0$ results in unrealistic overheating. To address this issue, Ghirardi \etal proposed a cutoff of $R_0=\unit{10^{-7}}{m}$, although without a fundamental justification \cite{ghirardi90}. Alternatively, Penrose proposed to determine $R_0$ by solving the Schrödinger-Newton equation \cite{penrose14}, which would make it system-dependent. Due to the lack of an unambiguous prescription for assigning its value, the cutoff $R_0$ is often considered as a free parameter of the DP model, which can be constrained by experimental measurements.

Inserting Eq.~\eqref{eq:MDensOp} into Eq.~\eqref{eq:Lform} we obtain \footnote{Here we used the fact that $\vert \vx-\vy \vert^{-1} = \int d^3 \vq e^{i(\vx-\vy)\cdot\vq/\hbar} / 2\pi^2 \hbar q^2$.}  
\begin{equation}\label{eq:DPcutoff}
     \mathcal{L}[\orho] = \dfrac{G}{2\pi^2\hbar^2} \sum_{j,l=1}^{N} m_j m_l \int \dfrac{d^3 \vq}{q^2} e^{-q^2 R_0^2/\hbar^2} \left( e^{i\vq\cdot\vrop_j/\hbar} \orho e^{-i\vq\cdot\vrop_l/\hbar} - \dfrac{1}{2} \lbrace e^{i\vq\cdot\vrop_j/\hbar} e^{-i\vq\cdot\vrop_l/\hbar}, \orho \rbrace \right)  \,. 
\end{equation}
In the following, we will compute this expression for the bulk acoustic wave mode of a crystal, and examine how DP decoherence depends on the system's parameters. Next, we will apply this result to analyze experimental data obtained from preparing an acoustic mode in a macroscopic superposition state, and derive the bound they put on $R_0$.

\clearpage
\newpage
\section{Testing the Diósi-Penrose model with bulk acoustic states}

We consider a collective acoustic excitation in the bulk condensed matter described by the mode function $\vu(\vr)$, which defines an effective oscillator mass as $m_{\mathrm{eff}} = \int d^3 \vr \, \varrho(\vr) u^2(\vr)$. In a quantum-mechanical description, this acoustic mode is associated to the bosonic ladder operator $\oa = (\ox + i \op)/2$, which involves dimensionless position and momentum quadrature operators satisfying $[\ox,\op]=i/2$. This allows us to write the position operator of the $j$th atom according to the displacement field as $\vrop_j = \vr_j + \vu(\vr_j)x_0 \ox$, where $x_0 = \sqrt{\hbar/m_{\mathrm{eff}}\omega}$ is related to the zero-point fluctuation as $x_0/\sqrt{2}$.

In order to compute Eq.~\eqref{eq:DPcutoff}, we use the fact that for small fluctuations around $\vr_j$ we can expand the exponentials to first order as $e^{i\vq\cdot\vrop_j/\hbar}=e^{i\vq\cdot\vr_j/\hbar}(1+i(\vq\cdot\vu(\vr_j))x_0 \ox/\hbar$. Furthermore, we consider a mode function $\vu(\vr) = u_x(\vr) \ve_x$.
The Liouvillian thus becomes the momentum diffusion equation
\begin{equation}
    \mathcal{L}[\orho] = \dfrac{G x_0^2}{2\pi^2\hbar^4} \int d^3 \, \vq \dfrac{q_x^2}{q^2} \vert \tilde{\varrho}(\vq) \vert^2 \left( \ox \orho \ox - \dfrac{1}{2} \lbrace \ox^2, \orho \rbrace \right) \equiv 2 \Gamma \mathcal{D}[\ox]\orho \;,
\end{equation}
where $\mathcal{D}[\ox]\orho = ( \ox \orho \ox - \lbrace \ox^2, \orho \rbrace/2 )$, and we introduced the Fourier transform of the coarse-grained mass density distribution
\begin{equation}
    \tilde{\varrho}(\vq) = \sum_{j=1}^N m_j u_x(\vr_j) e^{-q^2 R_0^2/2\hbar^2} e^{i\vq\cdot\vr_j/\hbar} \;.
\end{equation}
The diffusion rate can be calculated for a crystal with lattice constant $a$ and mean density $\overline{\varrho}$
\cite{StefanPRL}, and it reads (see Appendix A)
\begin{equation}\label{eq:GammaCrystal}
    \Gamma \approx \dfrac{G}{12\sqrt{\pi}\omega}\left(\dfrac{a}{R_0}\right)^3 \overline{\varrho} \;.
\end{equation}
Interestingly, note that this expression does not depend on the total mass of the system, but on its density \cite{diosiMassRes13}.

Taking into account the dynamics resulting from energy relaxation at rate $\gammadown$, and in the co-rotating frame of the oscillator, the time evolution of $\orho$ can be written after averaging over fast oscillating terms as $\dot{\orho}\approx(\Gamma+\gammadown)\mathcal{D}[\oa^\dagger]\orho + \Gamma \mathcal{D}[\oa]\orho$ \cite{macroPRL23}. 
This master equation admits a compact solution in the Wigner function representation, namely
\begin{equation}\label{eq:WEvolTh}
    W(X,P;t) = \dfrac{e^{\gammadown t}}{\pi S(t)} \int dX' dP' \; W(X'e^{\gammadown t/2},P'e^{\gammadown t/2};0) e^{-[(X-X')^2 + (P-P')^2]/S(t)} \;,
\end{equation}
where $S(t)=(2\Gamma/\gammadown +1)(1-e^{- \gammadown t})$. 

Note that the absence of a term $\gamma_\uparrow \mathcal{D}[\oa]\orho$ in the master equation corresponds to the conservative assumption of a zero-temperature environment, since by attributing any decoherence beyond that associated with relaxation to a gravity-related modification results in overestimating $\Gamma$. Within this assumption, and for $\Gamma\ll\gammadown$, this model predicts a population decay at rate $1/T_1\approx\gammadown$ \cite{macroPRL23}.

Given Wigner function measurements performed at times $t=\{0,t_1, ...\}$ and a value of $\gammadown$ obtained from an independent measurement of $T_1$, Bayesian parameter estimation based on Eq.~\eqref{eq:WEvolTh} enables us to determine the most conservative 5\% quantile of the diffusion rate $\Gamma$. Although smaller values of $\Gamma$ may also be consistent with the data, as the observed decoherence may also arise from unspecified classical noise sources, larger rates can be confidently excluded. Finally, using the inferred $\Gamma$ in Eq.~\eqref{eq:GammaCrystal} allows us to lower bound $R_0$.

\clearpage
\newpage

\begin{figure}[h]
  \centering
  \includegraphics[width=\textwidth]{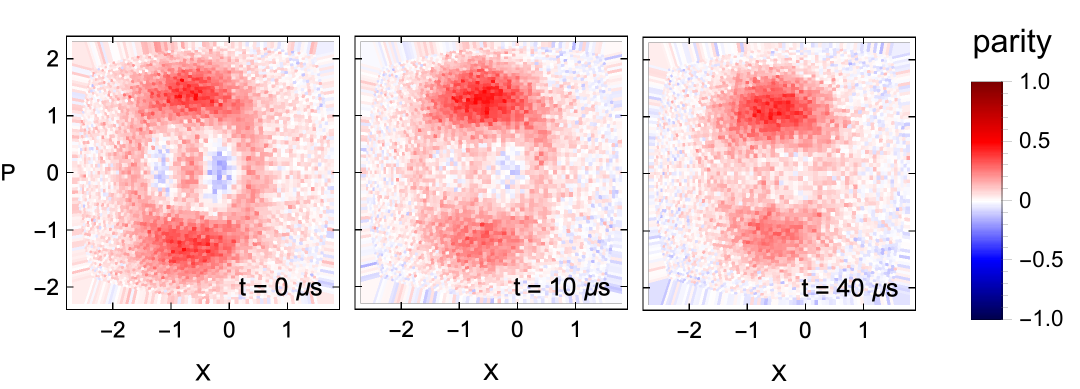}
  \caption{\textbf{Decoherence of a $\mathbf{16\,\mu g}$ Schrödinger cat state.} Measurements of the Wigner function taken at different times $t$ after having prepared the mechanical resonator in a cat state with average phonon number $\overline{n}=2.1$, reproduced from \cite{catSCI23}. The gradual disappearance of negative Wigner function values, as well as the shrinking of the state towards the phase-space origin, are the effect of decoherence.}
  \label{fig:figure2}
\end{figure}

\section{Outline of the experiment}
We outline here the experiment presented in Ref.~\cite{catSCI23}, where a high-overtone bulk-acoustic wave resonator (HBAR) was prepared in a Schrödinger cat state. This was achieved by coupling the HBAR to a superconducting qubit through a slab of piezoelectric material, Fig.~\ref{fig:figure1}. By exploiting the coupled system described by the Jaynes-Cummings Hamiltonian, we were able to utilize the toolbox of circuit quantum acoustodynamics to control the acoustic mode of the HBAR, which is localized in the bulk of a sapphire substrate \cite{vonLupke22}.

The experimental sequence is as follows. Initially, as the device is cooled down to millikelvin temperatures, both the qubit and the phonon mode are in the ground state with a negligible ($< 2\%$) thermal population. Then, after preparing the qubit in the excited state and the phonon mode in the coherent state $\vert \alpha\rangle = e^{-\vert\alpha\vert^2/2} \sum_n \alpha^n/\sqrt{n!} \vert n\rangle$, the two systems are allowed to evolve on resonance. Through the coupling with the qubit, the phonon coherent state undergoes a non-linear evolution, resulting in a state that resembles the coherent state superposition $\vert\alpha\rangle+\vert-\alpha\rangle$ after a specific evolution time \cite{Gea91}. This state, known as a Schrödinger cat state, has a mean phonon number of $\overline{n}=\vert\alpha\vert^2$. After state preparation, the qubit is decoupled from the resonator, which is left evolve freely for a time $t$. The Wigner function of the final state is then measured at different points, by using the qubit to perform displaced parity measurements \cite{vonLupke22}. 

The state under consideration has an average phonon number $\overline{n}=2.1$ at $t=0$. 
Considering the oscillator frequency $\omega=2\pi\,\unit{5.961}{GHz}$, this value corresponds to having the effective mass $m_\text{eff}=\unit{16.2}{\mu g}$ in a spatial superposition that is delocalized by $6.4$ times the zero-point fluctuation $\sqrt{\hbar/2m_\text{eff}\omega}=\unit{1.9 \cdot 10^{-18}}{m}$ \cite{catSCI23}.

Figure \ref{fig:figure2} shows the Wigner functions measured at times $t=\{0,10,40\}\unit{}{\mu s}$ after state preparation. Note the shrinking of the state towards the phase space origin as a result of energy relaxation, and the disappearance of negative (blue) regions over time as a result of decoherence, which turns the Wigner function into a distribution that admits a classical phase-space interpretation. The longer the initial Wigner negativity remains, the more strongly the experiment falsifies gravity-related modifications of quantum mechanics, such as the DP model. In the following section we are going to investigate the bound put on the free parameter $R_0$ of Eq.~\eqref{eq:DPcutoff} by the measurements shown in Fig.~\ref{fig:figure2}.

\clearpage
\newpage

\begin{figure}[h]
  \centering
  \includegraphics[width=\textwidth]{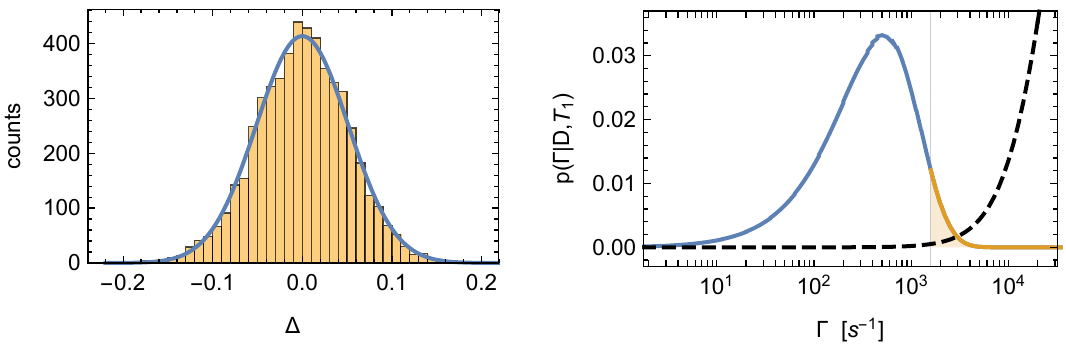}
  \caption{\textbf{Bayesian inference of the diffusion rate.} a) Histogram of the difference $\Delta$ between the Wigner function measurements taken at $t=0$, Fig.~\ref{fig:figure2}, and the corresponding reconstructed state, Fig.~\ref{fig:figure1}c. A Gaussian fit (blue line) gives a standard deviation for the distribution of $s=0.051$.
  b) Solid line: posterior distribution for the diffusion rate obtained from Bayesian inference with the data show in Fig.~\ref{fig:figure2}. The dashed line indicates the Jeffreys prior used, while the vertical line and the yellow region highlights the the lower $5\%$ quantile of the posterior distribution.
  \label{fig:figure3}}
\end{figure}

\section{Data analysis}

We first measure the Wigner function of the state prepared at time $t=0$, and run a maximum likelihood reconstruction algorithm to ensure that we obtain a physical state \cite{catSCI23}. The result is shown in Fig.~\ref{fig:figure1}c. This state can now be propagated in time through Eq.~\eqref{eq:WEvolTh}, by numerical integration. By comparing the experimental measurements of the state decoherence dynamics, Fig.~\ref{fig:figure2}, with the numerical predictions obtained for different $\Gamma$, we can use Bayesian inference to find the most likely values for the diffusion rate.

The posterior distribution for $\Gamma$, given the experimental data $\mathbf{D}$ and the independently measured decay time $T_1=\unit{84}{\mu s}$, is given by Bayes' theorem as $p(\Gamma\vert \mathbf{D}; T_1) \propto p(\mathbf{D}\vert\Gamma;T_1)p(\Gamma\vert T_1)$. Here, $\mathbf{D}=\{d_{ij}(t)\}_{ij}$ is the list of Wigner function points measured at time $t=\{10,40\}\unit{}{\mu s}$ (\ie the pixels in Fig.~\ref{fig:figure2}), $p(\mathbf{D}\vert\Gamma;T_1)$ is the likelihood for $\mathbf{D}$, and $p(\Gamma\vert T_1)$ the prior. We calculate the likelihood from evolving numerically the reconstructed state at time $t=0$ according to Eq.~\eqref{eq:WEvolTh}, and then evaluating the resulting Wigner function at the coordinates of the measured pixels. The resulting values are then convoluted with a Gaussian distribution of standard deviation $s$ to take into account measurement noise, which gives
\begin{equation}
    p(\mathbf{D}\vert\Gamma;T_1) = \prod_{i,j,t} \dfrac{1}{\sqrt{2\pi s^2}} \exp\left[ - \left\vert d_{ij}(t) - W(X_i,P_j;t) \right\vert^2 / 2 s^2 \right] \;.
\end{equation}
The value of $s=0.051$ is measured independently from the difference between the pixels measured at $t=0$ and the corresponding reconstructed state, see Fig.~\ref{fig:figure3}a, and it agrees with the noise level of the background pixels.
As prior distribution we chose Jeffreys prior, which is maximally objective in comparing different experiments \cite{BjornBayes}. This can be calculated as
\begin{equation}
    p(\Gamma\vert T_1) \propto \sqrt{ \left\langle \left( \dfrac{\partial}{\partial\Gamma}\log p(\mathbf{D}\vert\Gamma;T_1) \right)^2 \right\rangle_{\mathbf{D}} } \;,
\end{equation}
where $\langle\cdot\rangle_{\mathbf{D}}$ is the average over all measurement results.

The posterior distribution $p(\Gamma\vert \mathbf{D}; T_1)$ obtained from Bayesian inference with the data show in Fig.~\ref{fig:figure2} is plotted in Fig.~\ref{fig:figure3}b, where we marked the upper $5\%$ quantile for $\Gamma$ by a vertical line. We obtain the threshold value $\Gamma^\ast = \unit{1.4\cdot 10^3}{s^{-1}}$, such that larger rates are ruled out with $95\%$ confidence. Finally, insering this number into Eq.~\eqref{eq:GammaCrystal} gives
\begin{equation}\label{boundR0}
    R_0 \geq 6.2 \cdot \unit{10^{-17}}{m} \;.
\end{equation}
In the above, we used the density of sapphire $\overline{\varrho}=\unit{3.98}{g/cm^3}$, and an effective lattice constant $a=\unit{1}{\mu m}$.

\section{Discussion}
We have used the experimental measurements on the decoherence of a mechanical Schrödinger cat state presented in Ref.~\cite{catSCI23} to test the Diósi-Penrose model. This model relates the wavefunction collapse mechanisms to gravity, and it includes a cutoff parameter $R_0$ that determines the spatial resolution of the collapse field. From our data analysis, we obtain the lower bound $R_0 \gtrsim \unit{10^{-17}}{m}$. This is hundred times smaller than the bound $R_0 \gtrsim \unit{10^{-15}}{m}$ expected from the argument that in order to resolve length scales smaller than the nucleons' size a relativistic theory might be needed. To obtain such a bound with a HBAR device we would need to measure a diffusion rate $\Gamma \lesssim \unit{0.1}{s}$, which requires exceptionally long phonon coherence times in the order of seconds. In addition, Eq.~\eqref{eq:GammaCrystal} shows the advantage of increasing the bulk material density and of decreasing the acoustic mode frequency. Further investigations in this direction could also explore the possibility that gravity-related collapse of acoustic modes in solids depends on their wavelength \cite{diosiBulk}.

Let us remember that a bound on $R_0$ can also be obtained from a classical measurement of the system's equilibrium temperature \cite{ghirardi90}. In fact, the diffusion dynamics combined with energy relaxation predicts a steady state energy that can be computed from Eq.~\eqref{eq:WEvolTh} to be $E=\hbar\omega(1+2\Gamma T_1)/2$. Therefore, a bound on $\Gamma$ can be obtained just by measuring the steady state phonon population in the system, which in our experiment is $\overline{n}=1.6\%$ \cite{macroPRL23}. Without taking into account other known decoherence processes or finite temperature of the environment that can be characterized by independent measurements, we can identify $\overline{n}=\Gamma T_1$, giving $\Gamma=\unit{1.9\cdot 10^2}{s^{-1}}$. The resulting bound on the DP model cutoff parameter is then $R_0\geq \unit{1.2\cdot 10^{-16}}{m}$. While more stringent than the bound Eq.~\eqref{boundR0}, let us emphasize that measuring the equilibrium temperature is a classical measurement that do not require a quantum state. For this reason, there is still a conceptual difference in Eq.~\eqref{boundR0}, as it is obtained from directly observing the decoherence of a superposition.

As an outlook, it would be interesting to see if measurements like the one reported in Ref.~\cite{catSCI23} could be used to test gravitational \cite{bassiGrav17,BassiGrav20} or gravity-related \cite{seven22} decoherence models, as well as nonlinear extensions of the Schrödinger equation \cite{diosi84,MacroQM13}. This could shed light on the interplay between quantum mechanics and gravity, and give us a better understanding of the fundamental nature of space-time.

\section*{Acknowledgements}
I would like to express my gratitude to Lajos Diósi and Roger Penrose for their invaluable discussions and insights on the topic. Moreover, I thank Stefan Nimmrichter, Bj\"orn Schrinski and Yiwen Chu for their valuable contributions in discussing the data analysis and providing feedback on the manuscript. This work was supported by The Branco Weiss Fellowship -- Society in Science, administered by the ETH Z\"{u}rich.

\clearpage
\newpage

\bibliographystyle{quantum}
\bibliography{testDP}

\onecolumn
\appendix

\section{DP diffusion rate for a lattice}
Our goal is to calculate the diffusion rate
\begin{equation}\label{eq:SI_Gamma}
    \Gamma = \dfrac{G x_0^2}{4\pi^2\hbar^4} \int d^3 \vq \dfrac{q_x^2}{q^2}  \left\vert \tilde{\varrho}(\vq) \right\vert^2 \;,
\end{equation}
where $\tilde{\varrho}(\vq)$ is the Fourier transform of the mass density distribution of the considered mode.
Following Ref.~\cite{StefanPRL}, for cubic and monoatomic crystal of lattice constant $a$ we have
\begin{equation}
    \varrho(\vr) = \sum_{j,k,l=-\infty}^\infty u_x(\vr) \varrho_A(\vr-\mathbf{R}_{jkl}) \;,
\end{equation}
with $\mathbf{R}_{jkl}=a(j,k,l)$ a lattice vector, $u_x(\vr)$ the function specifying the mode profile, and $\varrho_A(\vr)=m_A \exp(-r^2/2R_0^2)/(2\pi R_0^2)^{3/2}$ the coarse-grained mass density distribution of each single atom (excluding the electronic contribution). Its Fourier transform is
\begin{equation}
    \Tilde{\varrho}(\vq) = \dfrac{1}{a^3} \sum_{j,k,l=-\infty}^\infty \Tilde{u}_x(\vq - \mathbf{Q}_{jkl}) \Tilde{\varrho}_A(\mathbf{Q}_{jkl}) \,
\end{equation}
with $\mathbf{Q}_{jkl}=2\pi\hbar(j,k,l)/a$ a reciprocal lattice vector, $\Tilde{\varrho}_A(\vq)=m_A \exp(-q^2 R_0^2/\hbar^2)$, and $\Tilde{u}_x(\vq)$ the Fourier transform of $u_x(\vr)$. Considering the latter to be a bulk acoustic wave in the GHz frequency range, note that in position space $u_x(\vr)$ varies over $\sim\unit{10^{-6}}{m}$, the lattice over $a\sim\unit{10^{-9}}{m}$, while $\varrho_A(\vr)$ over $\sim\unit{10^{-15}}{m}$. Given this hierarchy of length scales, we can conclude that in momentum space the function $\Tilde{u}_x(\vq)$ will be sharply peaked around $\vq$, which allows us to write
\begin{equation}
    \vert \Tilde{\varrho}(\vq) \vert^2 \approx \dfrac{1}{a^6} \sum_{jkl} \vert \Tilde{u}_x(\mathbf{q}-\mathbf{Q}_{jkl}) \Tilde{\varrho}_A(\mathbf{Q}_{jkl}) \vert^2 \;,
\end{equation}
as well as to approximate the diffusion rate with
\begin{align}
    \Gamma &\approx \dfrac{G x_0^2}{4\pi^2\hbar^4} \dfrac{1}{a^6} \sum_{jkl} \vert \Tilde{\varrho}_A(\mathbf{Q}_{jkl}) \vert^2 \int d^3 \vq \dfrac{q_x^2}{q^2} \vert \Tilde{u}_x(\mathbf{q}-\mathbf{Q}_{jkl}) \vert^2 \notag\\
    &\approx \dfrac{G x_0^2}{4\pi^2\hbar^4} \dfrac{1}{a^6} \sum_{jkl} \vert \Tilde{\varrho}_A(\mathbf{Q}_{jkl}) \vert^2 \dfrac{j^2}{j^2+k^2+l^2} \int d^3 \vq \vert \Tilde{u}_x(\mathbf{q}) \vert^2 \;.
\end{align}
Using Parseval's theorem we have $\int d^3 \vq \vert \Tilde{u}_x(\mathbf{q}) \vert^2 = (2\pi)^3 \int d^3 \vr \vert u_x(\vr) \vert^2 = (2\pi)^3 V_\text{eff}$. Furthermore, since in momentum space $\Tilde{\varrho}_A$ is much broader than the reciprocal lattice, we can now take the continuum approximation by replacing $\sum_{jkl}\rightarrow (a/2\pi)^3 \int d^3\vq$. This gives 
\begin{align}
    \Gamma &\approx \dfrac{G x_0^2}{4\pi^2\hbar^4} \dfrac{1}{a^6} (2\pi)^3 V_\text{eff} \left(\dfrac{a}{2\pi}\right)^3 \int d^3\vq \dfrac{q_x^2}{q^2} \vert \Tilde{\varrho}_A(\vq) \vert^2  \notag\\
    &\approx \dfrac{G x_0^2}{4\pi^2\hbar^4} \dfrac{1}{a^6} (2\pi)^3 V_\text{eff} \left(\dfrac{a}{2\pi}\right)^3 \left( m_A^2 \dfrac{\pi^{3/2}\hbar^3}{3 R_0^3} \right) = \dfrac{G x_0^2 m_A^2 V_\text{eff}}{12 \sqrt{\pi} \hbar a^3 R_0^3} \,.
\end{align}
Since the effective mass is $m_\text{eff}=\overline{\varrho} V_\text{eff}$, with $\overline{\varrho}=m_A/a^3$ the average density, we can write
\begin{equation}
    \Gamma \approx \dfrac{G x_0^2}{12\sqrt{\pi}\hbar}\left(\dfrac{a}{R_0}\right)^3 \overline{\varrho} \, m_\text{eff} \;.
\end{equation}

\end{document}